\documentclass[preprint]{aastex}

\usepackage[normalem]{ulem}
\usepackage{xcolor}
\usepackage{amsmath}

\newcommand{\eg}{{e.g.,}}

\newcommand{\txs}{TXS\,2332+154}
\newcommand{\sersic}{S\'{e}rsic}

\begin{document}

\title{A Compact Group of Galaxies at \boldmath$z = 2.48$ hosting an AGN-Driven Outflow\footnotemark[1]}

\footnotetext[1]{ 
Some of the data presented herein were obtained at the W.M. Keck Observatory, 
which is operated as a scientific partnership among the California Institute of Technology, the 
University of California and the National Aeronautics and Space Administration. The 
Observatory was made possible by the generous financial support of the W.M. Keck 
Foundation. The work is also based, in part, on data collected at Subaru Telescope, which is operated by the National Astronomical Observatory of Japan,
and on observations obtained at the Gemini Observatory, which is operated by the 
Association of Universities for Research in Astronomy, Inc., under a cooperative agreement 
with the NSF on behalf of the Gemini partnership: the National Science Foundation 
(United States), the National Research Council (Canada), CONICYT (Chile), the Australian 
Research Council (Australia), Minist\'{e}rio da Ci\^{e}ncia, Tecnologia e Inova\c{c}\~{a}o 
(Brazil) and Ministerio de Ciencia, Tecnolog\'{i}a e Innovaci\'{o}n Productiva (Argentina).}

\author{Hsin-Yi Shih} 
\affil{Gemini Observatory, 670 N Aohoku Pl., Hilo, HI, 96720}
\email{jshih@gemini.edu}
\and
\author{Alan Stockton}
\affil{Institute for Astronomy, University of Hawaii, 2680 Woodlawn
 Drive, Honolulu, HI 96822}
 \email{stockton@ifa.hawaii.edu}

\begin{abstract}

We present observations of a remarkable compact group of galaxies at $z = 2.48$. Four galaxies, all within 40 kpc of each other, surround a powerful high redshift radio source. This group comprises two compact red passive galaxies and a pair of merging galaxies. One of the red galaxies, with an apparent stellar mass of $3.6\times10^{11} M_{\odot}$ and an effective radius of 470 pc, is one of the most extreme examples of a massive quiescent compact galaxy found so far. One of the pair of merging galaxies hosts the AGN producing the large powerful radio structure. The merger is massive and enriched, consistent with the mass-metallicity relation expected at this redshift. Close to the merging nuclei, the emission lines exhibit broad and asymmetric profiles that suggest outflows powered either by a very young expanding radio jet or by AGN radiation. At $\gtrsim 50$ kpc from the system, we found a fainter extended-emission region that may be a part of a radio jet-driven outflow. 

\end{abstract}


\section{Introduction}

While carrying out a search for massive, passively evolving galaxies in high-redshift radio-source fields, we have come across a compact group of massive galaxies in the field of the radio source TXS2332+154, at $z=2.48$ \citep{de-01}.
Briefly, we have been targeting radio-source fields (1) because they likely represent overdense regions at high redshifts \citep[][and references therein]{wyl13}, and (2) because, knowing the redshift of the radio source, we can target redshifts for which standard photometric diagnostics can efficiently distinguish old stellar populations from heavily reddened star-forming galaxies.  In particular, the strong 4000 \AA\ break prominent in old populations falls between the $J$ and $H$ bands for $z\sim2.5$ \citep{sto04,sto07,sto08}. For radio-source fields at $z\sim2.5$, we look for galaxies with $K'<22$, $H\!-\!K'\approx0.6$, and $J\!-\!H\approx1.4$ (AB system), which would be consistent with a galaxy with an old stellar population with a mass $\gtrsim2\times10^{11} M_\odot$. See \citet{sto08} for more details. Out of some 30 radio-source fields with $2.25<z<2.75$ and radio flux densities $>0.5$ Jy at 365 MHz that we have observed, \txs\ is the only one that shows a compact group of bright galaxies near the radio source.

We have given a brief preliminary account of the old compact galaxy in the TXS2332+154 group in \citet{sto14}. Here we follow up that mention by describing the nature of the group in more detail, including the presence of a likely AGN-driven outflow. We also discuss the structural parameters of the galaxies from laser-guide-star adaptive-optics  (LGSAO) imaging, the dominant stellar populations from photometry over the range from 0.7 $\mu$m to 8 $\mu$m (observed frame), spectroscopy of some of the members, the identification of the galaxy responsible for the radio source, evidence for an ionized gas outflow from the radio source, and the discovery that one of the galaxies originally supposed to be part of this group is actually a projected foreground interloper.

\section{Imaging, Photometry, and Morphologies}

\subsection{Ground-Based Imaging}\label{gb}
Our initial object selection observations were obtained with the CISCO IR camera \citep{mot02} on the Subaru 8.2 m telescope \citep{iye04}. These comprised $J$, $H$, and $K'$ imaging with a scale of 0.105\arcsec\ pixel$^{-1}$ and a field of 1.8\arcmin. The PSF FWHM for these was 0\farcs56, and the deepest imaging was reached in the $H$ band, with a 1 $\sigma$ surface-brightness limit of 22.4 arcsec$^{-2}$ (AB mag).
Subsequent shorter-wavelength observations were then obtained to constrain the rest-frame UV SED. These consisted of $i'$ imaging with the Gemini Multi-Object Spectrograph (GMOS) on the Gemini North telescope and $I$ imaging with the Low-Resolution Imaging Spectrograph \citep[LRIS;][]{oke95} on the Keck I telescope.

In addition to these ground-based observations obtained to define the spectral-energy distribution (SED), we also obtained laser-guide-star adaptive-optics (LGSAO) $K'$ imaging with the NIRC2 camera on the Keck II telescope to obtain high-resolution morphological information. The camera was used in its wide-field mode, giving a pixel scale of 40 mas ($=330$ pc at $z=2.48$) and a field of 40\arcsec. For the final combined image the PSF FWHM  was 81 mas, and the 1 $\sigma$ surface-brightness limit was 25.1 arcsec$^{-2}$ (AB mag).

Both the CISCO and the NIRC2 LGSAO images were processed using our standard IRAF IR reduction pipeline, which does an iterative removal of the sky background, flatfielding, and drizzling the individual frames onto a final grid. The only significant difference between the processing of the CISCO and NIRC2 images was that the NIRC2 images were drizzled onto a $2\times$ subsampled grid prior to further processing.

\subsection{Spitzer IRAC and MIPS Imaging}\label{spitz}

The range of acceptable population synthesis models can be greatly constrained with longer wavelength photometry. We obtained deep imaging of the \txs\ field with the Spitzer Space Telescope in Cycle 3, including all 4 Infrared Array Camera (IRAC) bands (FWHM  $\sim1\farcs7$--2\farcs1) and the Multiband Imaging Photometer (MIPS) 24 $\mu$m band (FWHM $\sim5\farcs5$).  For each of the IRAC bands, we obtained a total of 3600 s of integration, using a 36-point Realeaux dithering pattern.  The total MIPS 24 $\mu$m integration for the field was 5400 s.

Starting from the IRAC basic calibrated data (BCD) produced by the pipeline, we used MOPEX with the ``drizzle'' option to combine the dithered images onto a $4\times$ oversampled grid in a way that would preserve as much resolution as possible in the final image for each band.  For the MIPS image, we simply used the final post-BCD image produced by the pipeline, confining additional processing to removing structure in the sky background.

\subsection{A Tight Group of Massive Galaxies}

\begin{figure*}[h!]
\epsscale{1.0}
\includegraphics[trim={0 5cm 0 2cm}, width=7.in]{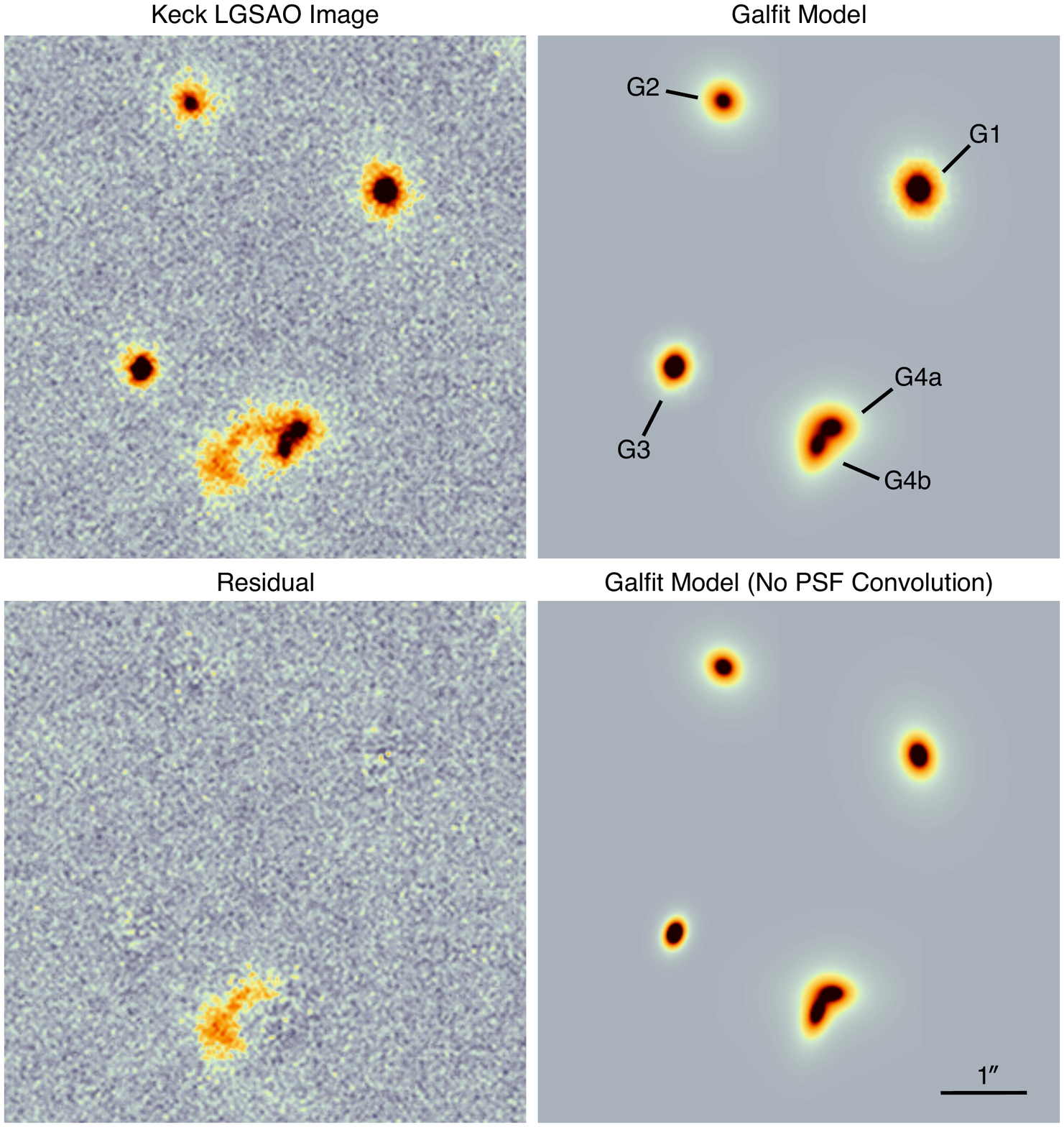}
\caption{Keck II LGSAO image of the TXS\,2332+154 group, with {\sc Galfit} models of each of the 5 galaxies. The upper-left panel shows the observed image; the upper-right panel shows the best {\sc Galfit} model, where all 5 galaxies are modeled simultaneously, after masking the region of the tidal tail from G2; the lower-left panel shows the residual from subtracting the model from the image; and the lower-right panel shows the model without convolving with the PSF. The PSF has a FWHM of 81 mas.}\label{galfit}
\end{figure*}

Figure \ref{galfit} shows our Keck LGSAO image of the \txs\ group, along with {\sc Galfit} \citep{pen02,pen10} modeling of the 5 galaxies, which allows us to estimate structural parameters for each. We have tested two approaches for the model fits: (1) fitting all galaxies simultaneously and (2) extracting regions around each galaxy and fitting them separately. Our final results were carried out with the latter approach, since it seemed to give better control over fitting the local sky background, which was modeled and subtracted. No constraints were used in the fitting, but the tidal tail of G4 was masked.  We used a star about 12\arcsec\ north of the group as the PSF reference; tests from a globular cluster field obtained on the same night indicate that the PSF was nearly uniform over separations of $\lesssim25\arcsec$. As mentioned above, the FWHM of the PSF star was 81 mas. In order to minimize noise in the wings of the profile, we fit a double Moffat profile to the star image (again using {\sc Galfit}). Within a radius of 0\farcs24, corresponding to an isophote of $\sim1$\% of the peak, the PSF was taken to be the profile of the star itself; for the wings beyond this radius, we used the model. We fitted \sersic\ profiles to the LGSAO images. From these fits, we can estimate the \sersic\ index, the effective radius $r_e$, and the axial ratio $b/a$.  These parameters are given in Table~\ref{table1}.

The uncertainties for these quantities derived from the fits are difficult to estimate. {\sc Galfit} does provide a purely statistical estimate of the uncertainties, but these are almost certainly unrealistically low in our case. We have resorted to a rough procedure that we believe will give a more realistic conservative estimate of these uncertainties. The most important parameters are the effective radius and the \sersic\ index. For these, we take the optimum solution found by {\sc Galfit} for each galaxy and run additional fits with each of these parameters fixed at various offsets from the optimum-fit value, allowing all other parameters to adjust. We exam the residuals and attempt to determine the point at which they become clearly unacceptable; we assume that this corresponds roughly to a $2 \sigma$ uncertainty in the associated parameter. The $1 \sigma$ uncertainties given in Table~\ref{table1} are based on this procedure.

\begin{deluxetable}{cccclcl}
\tablewidth{0pt}
\tablecaption{Structural Parameters for Galaxies in the TXS\,2332+154 Field}
\tablehead{
\colhead{Galaxy} & \colhead{$z$} & \colhead{$R_e$} & \colhead{$n$} & \colhead{$b/a$} & &\colhead{$M_{\rm stel}$}\\
& & (kpc) & &  & & \phantom{cel}($M_{\odot}$)
}
\startdata
G1\phantom{a} & $1.061\pm0.002$ & $2.16\substack{+1.08 \\ -0.43}$ & $5.96\substack{+2.33 \\ -1.04}$ & 0.75 & & \phn$3.8\times10^{10}$\tablenotemark{a} \vspace{1mm}\\
G2\phantom{a} & 2.48:$\phantom{0\pm0000}$ & $1.83\substack{+0.92 \\ -0.28}$ & $2.97\substack{+0.93 \\ -0.41}$ & 0.89 & & \phn$3.6\times10^{11}$ \vspace{1mm}\\
G3\phantom{a} & $2.482\pm0.001$ & $0.47\substack{+0.05 \\ -0.07}$ & $2.65\substack{+0.50 \\ -0.11}$ & 0.71 & & \phn$3.6\times10^{11}$\tablenotemark{b} \vspace{1mm}\\
G4a & $2.469\pm0.002$ & $2.98\substack{+1.79 \\ -0.52}$ & $4.56\substack{+1.25 \\ -0.74}$ & 0.69 & &  \vspace{1mm}\\
G4b & $2.469\pm0.002$ & $1.77\substack{+1.06 \\ -0.31}$ & $1.81\substack{+0.93 \\ -0.27}$ & 0.52 & \raisebox{1.7ex}[0pt]{\hspace{-5mm}\Large\}} & \raisebox{1.7ex}[0pt]{\phn$2.5\times10^{11}$\tablenotemark{c}}
\enddata
\tablenotetext{a}{G1 required a 2-component fit, one component of which is unresolved ($\lesssim100$ pc), comprising $\sim20$\% of the total flux at $K^{\prime}$. }
\tablenotetext{b}{The mass given is for the 1.9 Gyr model shown in Fig.~\ref{sed}. The mass for the 2.3 Gyr model would be $4.9\times10^{11} M_{\odot}$.}
\tablenotetext{c}{Combined mass of the two interacting galaxies, assuming the SED shown in Figure~\ref{sed} for the photometry corrected for emission-line contributions.}\label{table1}
\end{deluxetable}

\subsection{Spectral-Energy Distributions and Stellar Populations}\label{seds}

The \txs\ field presented a special challenge for photometry of the seeing-limited images, because we have 5 luminous galaxies all within a 5\arcsec\ circle. As can be seen in Figure~\ref{galfit}, except for a tidal tail from one of the galaxies, our models have very small residuals. In order to carry out accurate photometry for the seeing-limited images in each band, for each galaxy independently, we used its model derived from the LGSAO image, adjusting for pixel scale and rotating it to each of our images (optical, NIR, and Spitzer IRAC), using {\sc Galfit} to convolve the galaxy model with the appropriate PSF kernel and scale the result to minimize the residual when the model was subtracted from the observed image. In this way, we could generate the photometry from the individual scaled model images. Of course, for all but the LGSAO image, we could not separate the galaxies comprising G4, so we have had to treat them together for all properties except for the structural parameters derived from the LGSAO image.

\begin{figure*}
\epsscale{1.0}
\includegraphics[trim={0 10cm 0 2cm}, width=6.in]{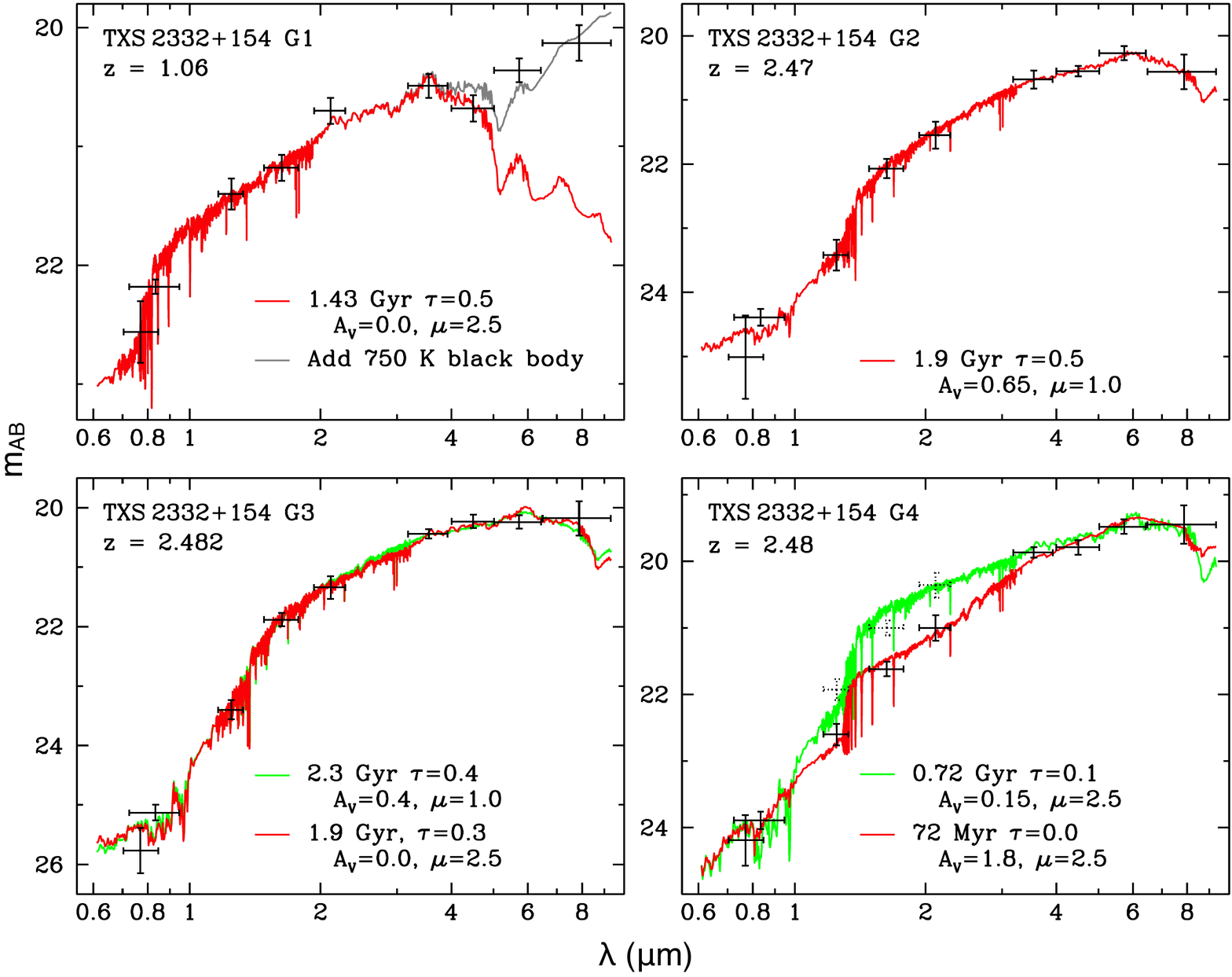}
\caption{SED photometry and models for the galaxies in the \txs\ group. Vertical bars on points indicate approximate 1 $\sigma$ uncertainties; horizontal bars indicate passbands between filter half-response points.  From left to right, the photometric bands are $i'$, $I_c$, $J$, $H$, $K'$, IRAC 3.6 $\mu$m, 4.5 $\mu$m, 5.7 $\mu$m, and 7.9 $\mu$m. The models shown are the best-fit exponentially decreasing star-formation models (with time constant $\tau$ given in Gyr) among those with metallicities $\mu=0.4$, 1.0, or 2.5 solar and \citet{cal00} reddening corresponding to extinctions $A_V$ ranging from 0 to 2.0. Note that the extremely compact passive galaxy G3 has two models that have similarly good fits, and that the long-wavelength upturn for G1 can be roughly fit by adding a 750 K black-body. For G4, the emission lines [\ion{O}{2}] $\lambda3727$, [\ion{O}{3}] $\lambda\lambda4959$,5007, and H$\alpha$ + [\ion{N}{2}] $\lambda\lambda6548$,6583 contribute significant portions of the flux in the $J$, $H$, and $K^{\prime}$ bands, respectively. For these bands, the original photometry is shown by dotted error bars, fitted by the green curve, while the photometry corrected by measurements of the line flux in our GNIRS spectra is shown by the solid error bars fitted by the red curve, which has a much younger age and much stronger reddening.}\label{sed}
\end{figure*}

\begin{figure*}
\epsscale{1.0}
\includegraphics[trim={0 10cm 0 2cm}, width=7.in]{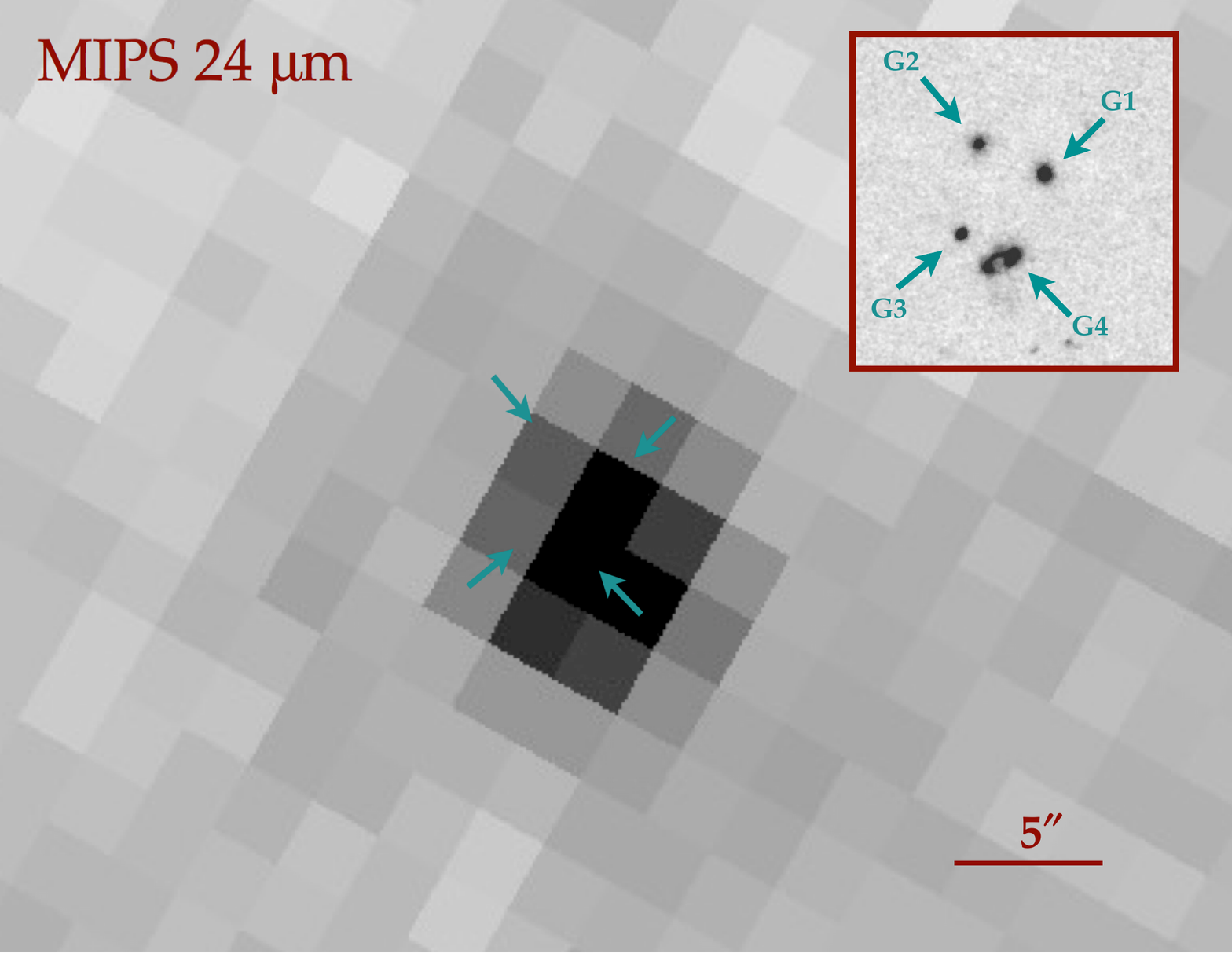}
\caption{MIPS 24 $\mu$m image of the \txs\ group. The inset shows the Keck II LGSAO image with arrows indicating the galaxies. In the MIPS image, the arrows are placed in the same position, where their position has been determined from astrometry of objects in the larger MIPS image. North is up and East is to the left.}\label{mips}
\end{figure*}

As described in \S\S~\ref{gb} and \ref{spitz}, we have a total of 10 photometric bands for the \txs\ field. Figure~\ref{sed} shows the photometry from all of these except the Spitzer MIPS 24 $\mu$m band, superposed on best-fitting Bruzual \& Charlot models \citep{bc03}.  The fitting was done with a modified version of the photometric redshift code {\em HyperZ} \citep{bol00}, allowing spectral-synthesis model, redshift, and \citet{cal00} reddening as free parameters. The model choices were restricted to instantaneous bursts and those with exponentially decreasing star-formation rates, each with solar, $2.5\times$ solar, and $0.4\times$ solar metallicities. All models had a \citet{cha03} initial mass function. An elongated detection at 24 $\mu$m, (see Figure~\ref{mips}) appears likely to have contributions, mainly from G4, but also probably from G1. For G4, this indicates the presence of warm dust reprocessing radiation produced by either an AGN (see \S\ \ref{radio}) or by star formation in this interacting pair. For G1, any 24 $\mu$m emission probably results mainly from the range of dust temperatures associated with the source of the apparent 750 K dust signature seen in the SED shown in Fig.~\ref{sed}. If {\em only} 750 K dust were present, the 24 $\mu$m emission would be about a factor of 2 below our 1 $\sigma$ detection threshold, so our detection indicates a likely spread of dust temperatures to lower values.  No adequate SED fit for G1 could be found for the radio-source redshift of $z=2.48$, hinting that it is likely at a different redshift, as we later confirmed via spectroscopy (\S~\ref{specoptir}).

Our GNIRS spectroscopy of G4 (see \S\ \ref{emstruct}) shows that emission lines make major contributions to the total flux in the $J$, $H$, and $K^{\prime}$ bands. In fact, the emission-line strength in each of these bands is such that the total equivalent widths are similar to the filter widths, requiring corrections to the observed fluxes for the fitting of stellar population models, as shown in Figure~\ref{sed}. We have made our best estimates of these corrections, but there remain possible systematic uncertainties that can not be easily quantified. With these corrections, we find a dominant contribution from a young ($\lesssim100$ Myr), highly reddened ($A_V\sim1.8$) stellar population. While the exact values may be somewhat uncertain, this qualitative result of a dominant young stellar population with substantial reddening is likely to be robust against any reasonable uncertainties in the exact corrections for the emission lines.

\section{Spectroscopy: Extended Emission and Ionization Mechanisms}

\subsection{Optical and NIR Spectroscopy}\label{specoptir}
\begin{figure*}[t!]
\epsscale{1.0}
\includegraphics[trim={0 14cm 0 2cm}, width=7.in]{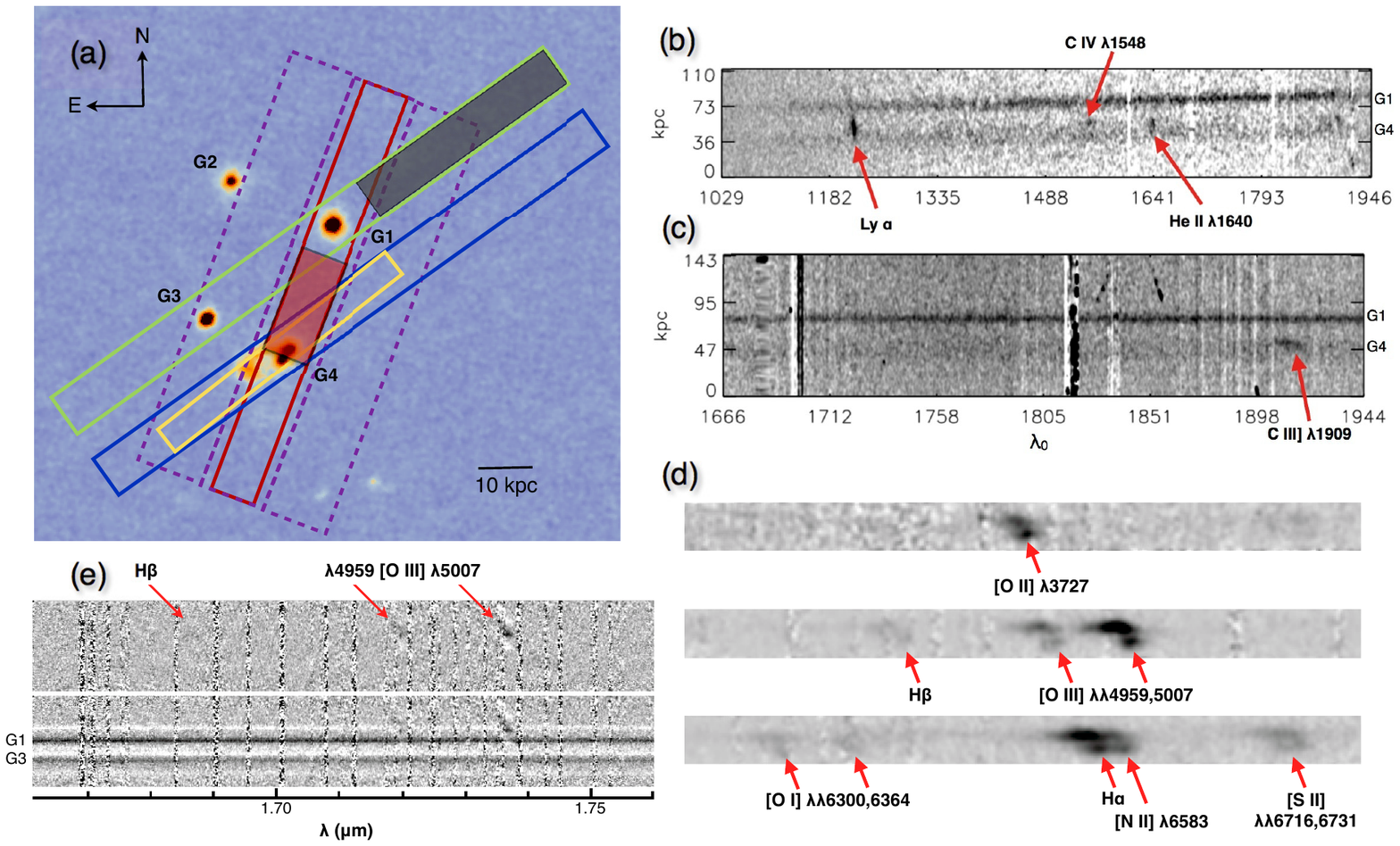}
\caption{Spectroscopy of the \txs\ field. ({\it a}) The position and slit width of the NIR and optical spectra overlaid on the NIRC2 LGSAO image. The MOSFIRE slit is shown in green with the gray shaded region indicating where [\ion{O}{3}] extended emission is detected. The GNIRS slit is shown in yellow. The slits for the optical spectra are shown as follows: purple for GMOS (including the offset positions), red for LRIS, and blue for the FOCAS. The red shaded region indicate where extended Ly $\alpha$ emission is detected with LRIS. The GMOS spectra are obtained with wider $1.5\arcsec$ slits, the GNIRS spectrum with $0.675\arcsec$ slit, and all others with $1\arcsec$ slits. (b) The 2-D LRIS spectrum at the red slit position shown in {\it a}, showing the strong Ly$\alpha$ feature extending out from G4 almost  to the G1 spectrum in the north, as well as faint detections of \ion{C}{4} $\lambda 1541$ and \ion{He}{2} $\lambda 1640$. ({\it c}) The 2-D GMOS spectrum at the central purple slit position shown in {\it a}, showing the broad [\ion{C}{3}] emission. ({\it d}) Parts of the GNIRS echellete spectrum showing regions around [\ion{O}{2}] $\lambda3727$, [\ion{O}{3}] $\lambda\lambda4959$,5007 + H$\beta$, and H$\alpha$ + [\ion{N}{2}] $\lambda6548$,6583. The upper trace in each spectrum is the central part of the G4 pair; the lower trace shows emission in the tidal tail. Note that [\ion{O}{2}] (and possibly [\ion{N}{2}]) are stronger in the tail, whereas [\ion{O}{3}] and H$\alpha$ are much stronger in the nuclear regions. ({\it e}) The 2-D MOSFIRE spectrum at the green slit position shown in {\it a}, showing only the H$\beta$---[\ion{O}{3}] region at $z\sim2.47$. The upper plot has had a model of the positive and negative continuum features produced by the MOSFIRE pipeline subtracted in order to show the emission more clearly.}\label{slitpos}
\end{figure*}

We have long-slit NIR (rest-frame optical) and optical (rest-frame UV) spectra of the compact group members at the slit positions shown in Figure \ref{slitpos}{\it a}. We obtained $H$-band spectra of two of the compact objects in the field (G1 and G3 in Figure \ref{galfit}) using MOSFIRE on Keck I, covering the 14680--18040 \AA\ region (rest-frame 4218--5184 \AA\ for $z = 2.48$). We also have a Gemini GNIRS cross-dispersed spectrum of the interacting pair (G4), covering 0.85--2.5 $\mu$m. We reduced the MOSFIRE spectrum using the MOSDRP pipeline (version 2014.06.10) developed by N. Konidaris and C. Steidel at Caltech, and the GNIRS spectrum with the Gemini IRAF reduction package. 

In the optical, we have three sets of spectroscopy from GMOS on Gemini North, FOCAS \citep{kas02} on Subaru, and LRIS \citep{mcc98,roc10} on Keck I, covering the rest-frame wavelength ranges of 1167--2376 \AA, 1135--1944 \AA, and 1029--3098 \AA\ respectively. The GMOS spectra include three positions: the center slit which passes through the interacting pair in the south (G4), and the blue compact object in the north (G1), offset $1.5\arcsec$ to the east, and offset $1.5\arcsec$ to the west, with position angle of $-21\arcdeg$. The LRIS spectrum covers the same central slit position. The three FOCAS slit positions have PA $= -50\arcdeg$, with the central slit passing through G4 and $1.5\arcsec$ offsets on each side. A summary of the spectral coverage of each galaxy is in Table 2, and Figure~\ref{slitpos}{\it a} shows the slit positions overlaid on the LGSAO image.  All optical spectra were reduced with IRAF using standard procedures for long-slit spectroscopy, including flat-fielding, wavelength calibration, sky subtraction, and flux calibration. 
 
In brief, the spectroscopy has confirmed that galaxy G3 and the interacting pair G4 are at $z\sim2.48$ but that galaxy G1 is a foreground interloper at $z=1.061\pm0.002$. The redshift for G3, $2.482 \pm 0.001$, was determined from H$\gamma$ and H$\beta$ absorption lines. The redshifts for G1 and G4 ($2.469 \pm 0.002$) were based on emission lines. Low redshift compact group members usually have low velocity dispersions, with most galaxies having velocity differences $< 500$ km s$^{-1}$ from the relative median velocity of each group \citep{hickson1997}. Using the relativistic Doppler formula, the redshifts of G3 and G4a and b are within $500$ km s$^{-1}$ of each other, and therefore they may be considered part of a compact group. Although we have no spectrum of G2, the SED of a 1.9 Gyr stellar population with a small amount of reddening at $z\sim2.47$ gives an excellent fit to the photometry, so there is little doubt that it also is a member of the group.

\begin{deluxetable}{cccccc}
\tablewidth{0pt}
\tablecaption{Spectral Coverage of Each Galaxy}
\tablehead{
\colhead{} & \colhead{LRIS} & \colhead{GMOS} & \colhead{GNIRS} &\colhead{FOCAS} &\colhead{MOSFIRE}\\
Slit width & $1\arcsec$\ & $1.5\arcsec$ & $0.675\arcsec$ & $1\arcsec$ & $1\arcsec$ \\
$R$ & 500  & 560 & 750 & 400 & 2560 \\
$\Delta \lambda$  & 3580-& 4060- & 8500- & 3950- & 14680- \\
 & 10780\AA & 8270\AA & 25000\AA & 6765\AA & 18040\AA \\
  }
\startdata
G1\phantom{a} & y & y & -- & -- & y \\
G2\phantom{a} & -- & -- & -- & -- & -- \\
G3\phantom{a} &-- & -- & -- & -- & y \\
G4 (a \& b) & y & y & y & y & -- \\
\enddata
\end{deluxetable}

\subsection{Emission Lines from the Interacting Pair}

\subsubsection{Emission-Line Structures}\label{emstruct}
The LRIS spectrum of interacting pair shows bright extended Ly$\alpha$ emission (Figure~\ref{slitpos}{\it b}), which spans a $\sim 20$ kpc region along the slit, almost extending out to the position of G1. The profile of the UV lines is best illustrated by the \ion{C}{3}] $\lambda 1909$ line in the GMOS spectrum (Figure \ref{slitpos}{\it c}), with a velocity dispersion FWHM = $590 \pm 50$ km s$^{-1}$ (all velocity dispersions and velocity differences mentioned in this paper have been calculated from differential redshifts using the relativistic Doppler formula).

The optical emission lines in our GNIRS spectrum (Figure \ref{slitpos}{\it d}), observed at a different PA from that of the UV lines ($-52\arcdeg$ for optical lines vs. $-21\arcdeg$ for UV lines) do not appear to be extended to the northwest. The [\ion{O}{3}] emission from the nuclear regions is roughly as broad as the \ion{C}{3}], but the higher signal-to-noise of this line also reveals a clear asymmetry. A single Gaussian provides a poor fit to the [\ion{O}{3}] profile, which requires two components. The dominant component of the best fit model has $FWHM = 255 \pm 15 $ km s$^{-1}$, and is accompanied by a fainter component that is slightly broader,  with $FWHM = 320 \pm 70$ km s$^{-1}$ and blueshifted by $-290 \pm 45$ km s$^{-1}$. The merger tail in the southeast has narrower emission with FWHM $= 210 \pm 15$ km s$^{-1}$ that is redshifted by $130 \pm 25$ km s$^{-1}$ relative to the dominant component of the nuclear emission. 

In addition to the bright extended emission lines close to the merger nuclei, we have detected fainter extended [\ion{O}{3}] $\lambda 5007$ emission on the north-west side of G1 in the MOSFIRE spectrum (Figure~\ref{slitpos}{\it e}). As we have eliminated G1 as a member of the $z = 2.48$ group, a hidden quasar in the merging pair G4---located $> 50$ kpc from the tip of the detected [\ion{O}{3}] emission---becomes the only likely ionizing source. The gas is more red-shifted in the southeast and more blue-shifted in the northwest, spanning a velocity range of $-140$ to $-240 \pm 20$ km s$^{-1}$ with respect to the dominant component of the nuclear emission. The gas velocity dispersion is relatively low, with $FWHM = 35 \pm 15$ km s$^{-1}$ over the detected range. Given the weakness of H$\beta$, we estimate the ratio [\ion{O}{3}]/H$\beta \gtrsim 5$, which is consistent with AGN ionization.

While the [\ion{O}{3}] extended emission is fairly luminous ($\sim 4\times10^{-16}$ erg s$^{-1}$ cm$^{-2}$), there is, at most, only a slight hint of extended Ly-$\alpha$ in our LRIS spectrum near this position, and we do not detect any other UV emission lines there. The 3 $\sigma$ detection limit for our LRIS spectrum at Ly-$\alpha$ on the north side of G1 is $\sim 3.3 \times 10^{-17}$ erg s$^{-1}$ cm$^{-2}$ arcsec$^{-2}$. We should emphasize, though, that the LRIS spectrum does not sample the exact region for which the MOSFIRE spectrum detected [\ion{O}{3}] emission (see Fig.~\ref{slitpos}{\it a}).




\subsubsection{Emission Line Ratios}\label{ratios}

\begin{figure*}[t!]
\centering
\includegraphics[trim={0 12cm 0 2cm}, width=7.in]{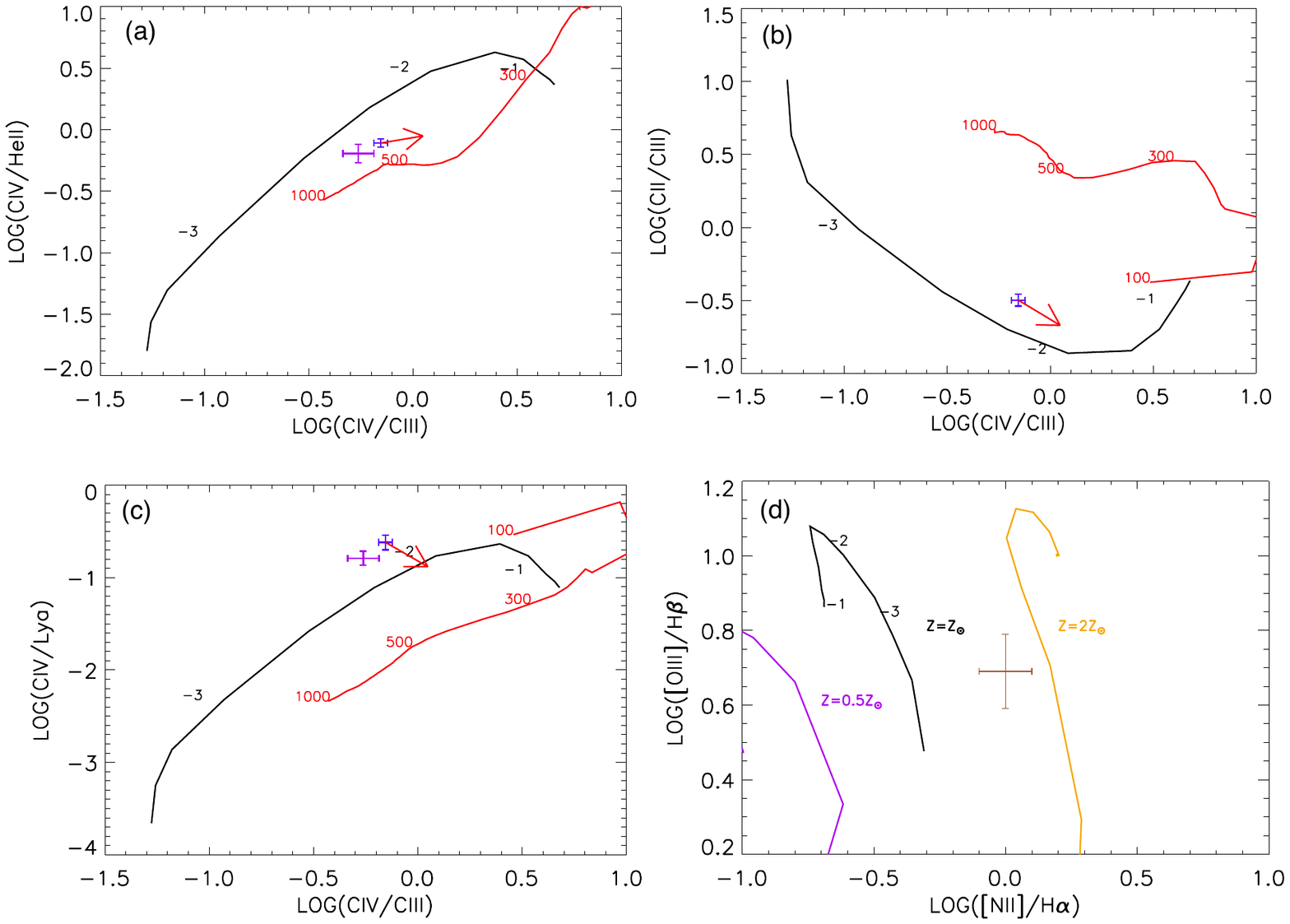}
\caption{Panels (a), (b), and (c): UV line ratio diagnostic diagrams, with dusty AGN ionization models (black), pure shock models (red), and the A$_{M/I}$ sequence from \citet{binette96}. The AGN models have solar metallicity, $n = 100$ cm$^{-3}$, and $\alpha = -1.2$,. The log($U$) values along the model are indicated by the black numbers. The pure shock models also have solar metallicity and $n = 100$ cm$^{-3}$, and the magnetic field is $B = 20$ $\mu$G. The red numbers along the shock models mark the shock velocities. The blue data points are the line ratios measured from the FOCAS spectra, and the purple data points are those measured from the GMOS spectra. Data measured from two difference set of spectra are in good agreement with each other, and fit the AGN ionization models well. Panel (d): Optical line ratios from the GNIRS spectrum plotted along with dusty AGN models with a range of metallicities, $\alpha = -1.2$, and $n = 100$ cm$^{-3}$. The data point falls between the models with solar and $2 \times$ solar metallicity. The corresponding shock model mostly falls off the plotted range, so it is not included in this panel.}\label{line_ratios}
\end{figure*}

\subsubsubsection{Line Ratio Measurements}
Relative emission-line strengths provide a powerful tool for diagnosing the excitation state and ionization mechanism of the gas. We measured the strength of the emission lines by fitting a single Gaussian component to each line using the MPFIT routine \citep{mpfit}. The line ratios, along with AGN and shock ionization models by \citet{groves04} and \citet{allen08}, are shown in Figure \ref{line_ratios}. The UV emission lines ratios are dominated by emission from the nuclear region. The optical emission-lines ratios, on the other hand, can only be accurately measured in tail region in the south-east. The [\ion{N}{2}]/H$\alpha$ ratio of the nuclear region is difficult to determine due to the blending of multiple velocity components. 

The merging galaxies are expected to have a significant amount of dust, which can affect the UV line ratios. While we do not have enough information to accurately determine the magnitude of the extinction in the emission-line region, we can examine how reddening may affect the line ratios by approximately de-reddening the spectra.  We use the reddening prescription of \citet{cal00}, which is suitable for dusty star-forming galaxies. Taking the best-fit extinction value from the SED fit, we de-reddened the spectra assuming an extinction of $A_v \sim 1.8$, and then re-measured the line ratios. We included a reddening vector in each diagnostic plot to show the result of de-reddening. Unless the actual extinction is much larger than $A_v = 1.8$, reddening does not change the best fit ionization model.  

Using Ly-$\alpha$ as a diagnostic line may seem unusual since it can be heavily affected by dust extinction enhanced by the additional path length due to resonant scattering. However, Ly-$\alpha$ lines from high-z radio galaxies are often not evenly absorbed across the whole line profile. Studies by \citet{vanojik97} and \citet{debreuck00} show that Ly$\alpha$ absorption tends to form deep troughs in the line profile and either cause the line to appear multi-peaked, or preferentially absorbed in the blue. The Ly-$\alpha$ from our target can be well fitted by a single Gaussian profile and therefore does not appear to be heavily absorbed. Furthermore, even if the Ly-$\alpha$ flux were to be reduced by up to $\sim 50\%$, it would not significantly affect the results from the diagnostic diagrams. 

\subsubsubsection{Ionization Mechanism and Abundance}

We expect the AGN to be a major contributor of the ionizing radiation. However, given the powerful radio jet associated with the system, shock ionizations due to jet-cloud interactions may also be significant. We compare the observed line ratios to two theoretical models. The first model, shown as black lines in Figure \ref{line_ratios}, is a dusty AGN model from \citet{groves04}, which takes the scattering, photoelectric heating, and radiation pressure due to dust into account. The input ionizing spectrum is a power law with index $\alpha = -1.2$. The second model, shown as red lines, is a shock model taken from \citet{allen08}. We consider a range of shock velocities with magnetic field of $B \sim 20$ $\mu$G. The shock models in \citet{allen08} include a wide range of magnetic fields to cover the extremes expected in the interstellar medium ($10^{-3} < B < 1000 \mu$G). We choose a value that lands around the middle of the extremes and likely represent a typical scenario. For both models, we assume that the gas has a density of $n = 100$ cm$^{-3}$ and solar metallicity. 

The dusty AGN model can produce line ratios that come reasonably close to the observed value. However, as shown in Figure \ref{line_ratios}, the AGN and shock models overlap around the area of interest for many of the line ratios. The [\ion{C}{2}]/\ion{C}{3}] plot is where the two models diverges the most, and the low [\ion{C}{2}]/\ion{C}{3}] value favors the AGN model over the shock model. The high [\ion{C}{4}]/Ly$\alpha$ ratio is also in better agreement with the AGN model. Therefore, while we cannot completely rule out the presence of shocks around the nuclear region, the AGN is likely the dominant source of ionization. 

We can make a rough metallicity estimate using the [\ion{N}{2}]/H$\alpha$ ratio, which is $0.0 \pm 0.12$ at the merger tail region. There are several ways to estimate the abundance with this ratio: (1) On a BPT diagram such as figure \ref{slitpos}{\it d}, the combination of the high [\ion{N}{2}]/H$\alpha$ and [\ion{O}{3}]/H$\beta$ ratios can only be produced by AGN ionization models with $Z \gtrsim Z_{\odot}$. (2) According to the empirical calibration by \citet{steidel14} (modified from the calibration by \citet{pp04}), the corresponding $12 + log(O/H) = 8.62 \pm 0.04$, which is close to the solar value of 8.69 \citep{asplund09}. (3) The SED fitting result, which gives $Z_{\odot} < Z < 2.5 Z_{\odot}$ as an acceptable range of metallicities, also points to a metallicity $\gtrsim Z_{\odot}$. The total mass of the merging pair is estimated to be $\sim 2.5 \times 10^{11} M_{\odot}$. If the metallicity of the tail region is representative of the entire merging system's metallicity, then this system is consistent with the mass-metallicity relation expected for $z \sim 2.3$ star-forming galaxies \citep[e.g.,][]{steidel14,kacprzak15}. Note that we cannot use the the [\ion{N}{2}]/H$\alpha$ ratio from the nuclear region because they are too broad to be reliably de-blended.


\section{Radio Source Identification}\label{radio}

The radio source has a flux density of 820 mJy at 365 MHz \citep{dou96} and a steep spectral index $\alpha_{365}^{1400}$ of 1.14 ($f_{\nu}\propto \nu^{-\alpha}$). The steep spectral index is not merely due to the object's high redshift. At lower observed frequencies, the spectral index between 74 MHz and 408 MHz, which translates to 258 MHz and 1420 MHz respectively in the rest frame, is 0.83. Since most of the radio emission comes from the unresolved core, this object can be considered a compact-steep-spectrum (CSS) source, which is most likely associated with young expanding radio jet(s) \citep{odea98}.

\begin{figure*}[t!]
\epsscale{1.0}
\includegraphics[trim={0 6cm 0 2cm}, width=7.in]{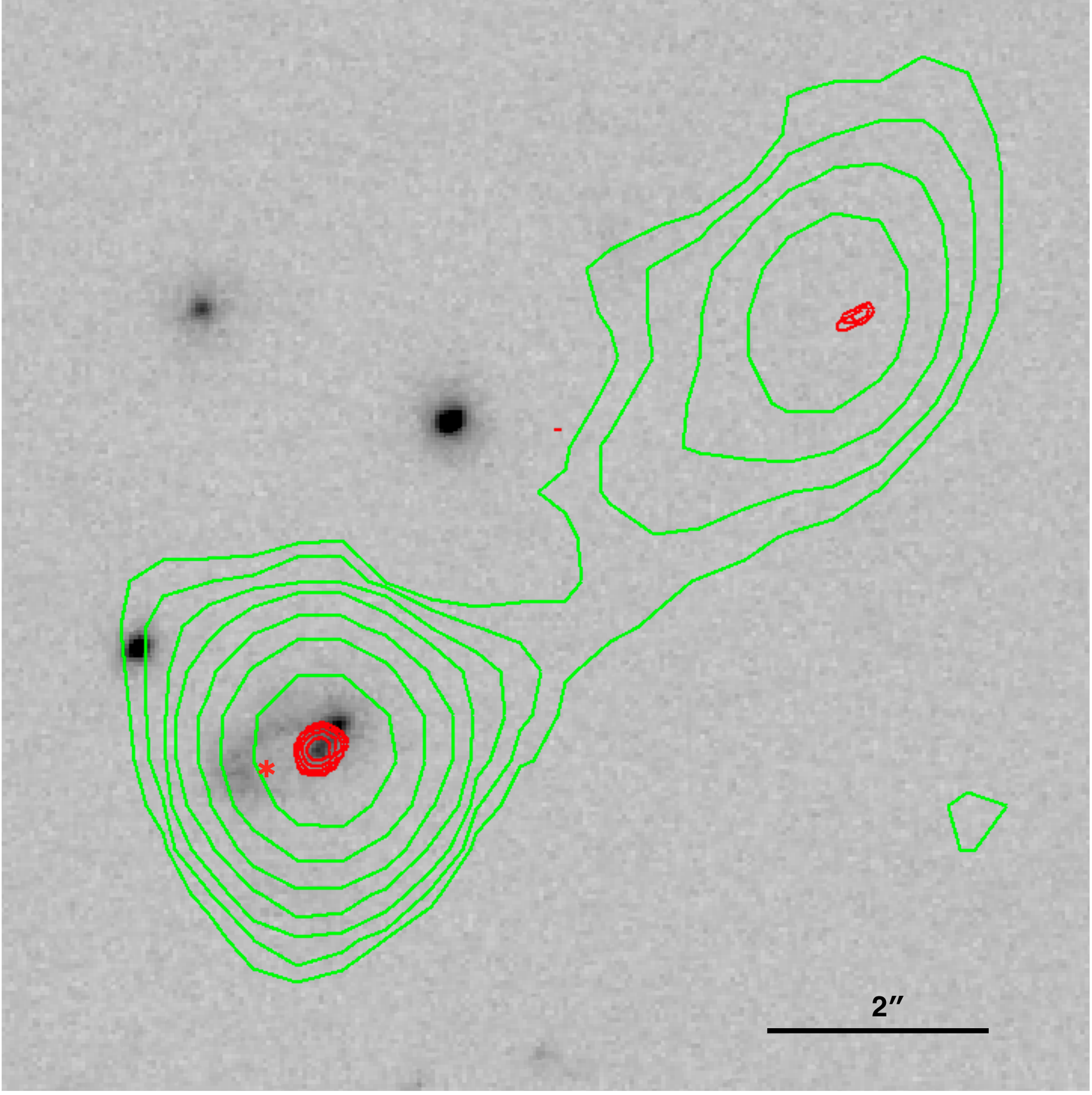}
\caption{VLA radio contours superposed on an image of the \txs\ group. The core component L-band (1.5 GHz) contours (green) and X-band (8.4 GHz) contours (red) have been centered on galaxy G4b, the closest galaxy to the nominal VLA position, which is indicated by the red asterisk. The L-band contour levels are at 1, 2, 4, 8, 16, 32, and 64 mJy per beam; those for the X band are 0.4, 0.5, 1, 2, 4, and 8 mJy per beam.}\label{vla}
\end{figure*}

The radio position of \citet{vil99} was close to G1, which they tentatively identified as the radio source. However, the Texas Survey position \citep{dou96} is closer to G4. In Figure~\ref{vla} we plot  archival VLA snapshot maps, which quite decisively favor one of the interacting galaxies comprising G4. This identification is reinforced by (1) the presence of extended emission at $z\sim2.47$ consistent with photoionization by a hidden quasar, which cannot be due to one in G1, at $z=1.06$, and (2) the apparent association of the strong Ly-$\alpha$ emission seen in Fig.~6 with G4.

\section{Discussion}

\subsection{The Galaxies in the Group}
All of the galaxies in this group (as well as the foreground galaxy G1) are quite compact; this is of course expected for massive galaxies at redshifts $>2$ \citep[\eg][]{vanD08}. We briefly discuss the properties of each of the galaxies:
\begin{itemize}
\item {\bf G1} we now know to be a foreground object, at $z=1.061$. As mentioned in \S\ref{seds}, we had a hint that this was the case even before obtaining our spectrum of it, as no reasonable SED at $z\sim2.48$ gave a good fit to the photometry. Even at the observed redshift, the galaxy is quite unusual, with its requirement of a two-component fit involving an extremely compact component and the need for an additional warm dust component to explain the upturn in the SED in the 2 long-wavelength IRAC bands and the apparent detection at 24 $\mu$m. Nevertheless, the fit to a 1.4-Gyr-old stellar population over the rest-frame 0.35 to 2.2 $\mu$m region is quite compelling. The apparent need for a component emitting black-body radiation at around 750 K indicates the likely presence of a hidden radio-quiet AGN in this galaxy, where the hot emitting dust is likely in the obscuring torus quite close to the AGN. The emission-line ratios we have available do not give unambiguous guidance on the ionizing source. The H$\beta$ line lies close to the edge of a strong airglow line in our LRIS spectrum, resulting in considerable uncertainty in the [\ion{O}{3}]/H$\beta$ ratio: we can only be confident that it is somewhere between 5 and 10. If near our lower limit, the ratio could be consistent with ionization either by star formation or an AGN; if near the upper limit, AGN photoionization is virtually certain. Our SED fit (Figure \ref{sed}) with a 1.4 Gyr population indicates that significant current star formation is unlikely, but we cannot completely rule out a level of star formation that could affect the emission lines that we see.
\item {\bf G2} is the galaxy for which we have the least data; but, as we mentioned earlier, the excellent fit of the stellar population model leaves little doubt that it is a member of the $z\sim2.48$ group. Although G2 has an identical stellar-population age to the most likely age for G3, and essentially the same estimated stellar mass, it is less compact, as indicated by its larger $R_e$ in Table~\ref{table1}. It also can clearly be seen to be more diffuse in Fig.~\ref{galfit}.
\item {\bf G3} is, as noted by \citet{sto14}, one of the most extreme examples of a compact quiescent galaxy at a high redshift. Packing $3.6\times10^{11} M_{\odot}$ of stars into a system with an $R_{\rm e}$ of only 470 pc indicates an extremely high stellar density. Assuming a dimension along the line of sight characterized by the circularized $R_{\rm e}$ in the plane of the sky, we can integrate the \sersic\ profile to estimate a mass density of $4.8\times10^{11} M_{\odot}$ kpc$^{-3}$ within $R_{\rm e}$ \citep[see, e.g.,][]{sar12}.
\item {\bf G4} is the merging pair of galaxies sporting an impressive tidal tail. One of these harbors a quasar, hidden from our vantage point, but betrayed by the compact-steep-spectrum (CSS) radio core as well as by the strong extended emission region. The radio-optical astrometry makes it most likely that the quasar is associated with G4b, but we cannot exclude G4a as the host. Our best estimate of the continuum SED (Figure~\ref{sed}), requiring a relatively young stellar population with a large amount of dust reddening, indicates that the luminosity in the rest-frame UV and optical is dominated by a recent massive starburst likely triggered by the merger in progress.
\end{itemize}

Isolated compact groups of galaxies are fairly common at low redshift \citep{hic82}, and a few similar groups have been found at redshifts up to $z\sim0.7$ \citep[e.g., MRC B0058$-$229, at $z=0.706$;][]{barr03}. \citet{amr08} have ventured to say that there are ``probably no (or a few) isolated CGs in the high $z$ universe,'' but in the \txs\ group, we have 4 massive galaxies (including the interacting pair) within a projected region $<40$ kpc across. Any similar galaxies in this field are at least 200 kpc away, and likely much farther, so this group clearly satisfies the criterion of being isolated.

\subsection{Outflow from the Merger Nuclei}

The nuclear [\ion{O}{3}] line profile has a blue wing which requires an additional Gaussian component, offset by $-290\pm45$ km s$^{-1}$ relative to the brighter component. Given that outflowing gas likely formed the extended Ly$-\alpha$ emission northwest of the merger, it may also be responsible for producing the fainter blue velocity component in the [\ion{O}{3}] profile. This situation is quite close to that found in the low-redshift compact-steep-spectrum quasar 3C\,48, which also involves an ongoing merger and a wide-solid-angle outflow apparently triggered by the initiation of the radio jet \citep{sto07b,shih14}. 

As mentioned in section \ref{radio}, there is likely a young emerging radio jet in this system, which can impart a large amount of energy to the surrounding gas and drive the nuclear outflow. Young radio sources at lower redshifts are known to drive outflows with complex velocity structures, often involving a broad and centrally concentrated component \citep[e.g.,][]{holt08}. 

Other possible energy sources that may contribute to driving the outflow are AGN radiation and supernovae from a startburst. AGN radiation heating is often invoked in galaxy evolution models to suppress star formation and further AGN activity \citep[e.g.,][]{hopkins05}. In a merger where the star-formation rate is expected to be high, energy injected by supernovae may also be significant. However, in powerful low-redshift CSS sources similar to \txs, the gas kinematics appear to be dominated by influences from radio jets \citep{shih13}. Therefore, the young radio jet(s) are most likely the dominant outflow-driving mechanism in G4 as well. 

\subsection{Extended Emission}
The Ly$\alpha$ emission, which extends $\sim 20$ kpc northwest of the merger, and the [\ion{O}{3}] feature at $\sim 50$ kpc are likely both formed by outflowing gas originating from one of the interacting pair of galaxies. The two red galaxies in the group are gas poor and not expected to produce outflows. 

Extended Ly-$\alpha$ nebulae are common around high redshift radio sources \citep[e.g.,][]{steidel00}. These extended blobs tend to show evidence of jet-gas interactions \citep[e.g.,][]{humphrey06}, and blue-shifted Ly-$\alpha$ absorption on top of the emission line suggesting outflows \citep{debreuck00}. Although the extended emission features in \txs\ are not aligned with the radio jet axis, they may still be jet-driven outflows. Wide-solid-angle blast waves associated with the launch of radio jets may be capable of driving outflows that do not appear to have any morphological correlation with the jets \citep[e.g.,][]{fu09}. Most extended emission-line nebulae around low-redshift radio sources have chaotic morphologies that do not necessarily correlate with the radio structure. If G4 is currently undergoing a starburst phase, the supernovae may also drive superwinds \citep[e.g.,][]{taniguchi00}. We will need more information of the star formation rate to determine whether supernovae can generate enough energy to drive outflows in this system. 

In the jet-driven superwind scenario, if the extended [\ion{O}{3}]-emitting gas is ejected from the current location of the merging pair then, assuming an average outflow speed of $\sim 500$ km s$^{-1}$, it will take at least $10^8$ years for the gas to travel the projected distance of 50 kpc. The typical lifetime for radio jets is estimated to be $\sim 10^7 - 10^8$ yrs \citep{shabala08}. Therefore, the tip of the detected [\ion{O}{3}] emission was likely ejected very close to the beginning of the radio activity. Alternatively, an earlier episode of feedback may be responsible. One possibly analogous local example of extended [\ion{O}{3}] emission around mergers is found in Mrk\,273 \citep{zaurin14}, where the emission-line gas extends up to $\sim 23$ kpc from the nucleus, and has similarly low velocity dispersion. \citeauthor{zaurin14} interpret this as an outflow from earlier stages of the merger that had cooled and slowed down.  For \txs, an earlier episode of feedback could have been triggered during a previous close passage. 

The faint extended [\ion{O}{3}] emission could also, in principle, result from infall, although this possibility seems much less likely in this case than an outflow. Two possible scenarios of infalling gas are: incoming cold streams \citep[e.g.,][]{geordt10}, or condensing halo gas \citep[e.g.,][]{haiman01}. In the former case, the collapsing gas will form a clumpy distribution around the galaxy, while the later case is expected to form a smoother distribution. Given the asymmetry of the extended [\ion{O}{3}] region, it would be more likely to be associated with cold streams if it were indeed to have been produced by infalling gas.

\section{Conclusion}

We have characterized the members of an extremely compact group of galaxies at $z = 2.48$ through laser-guided AO imaging and optical/near-IR photometry and spectroscopy. One of the putative members (G1) is a compact foreground interloper. Two red galaxies (G2 and G3) are both small in size and yet  fairly massive. G3, with an apparent stellar mass of $3.6\times10^{11} M_{\odot}$ and an effective radius of 470 pc, is certainly one of the most extreme examples of a massive compact quiescent galaxy identified so far.

One of the the two merging galaxies comprising G4 is clearly also the host for the AGN that powers the radio emission. The [\ion{O}{3}] line close to the nuclear region has two velocity components, offset by $290$ km s$^{-1}$, suggesting possible outflows powered by the AGN, either from the action of the powerful CSS jet on the surrounding medium or through radiative coupling with dust. This outflow may be responsible for the bright extended Ly$\alpha$ gas detected northwest of the galaxies. Farther from the merger, we detect fainter extended [\ion{O}{3}] emission reaching out to $> 50$ kpc, which is likely part of an outflow associated with the more extended component of the radio jet. 

Compact groups like that associated with \txs\ appear to be rare at $z\sim2.5$. Two of the galaxies (G2 and G3) in this group already seem to have formed virtually all of their stars between $z\sim5$ and $z\sim7$. The remaining two galaxies (G4a and G4b) have likely just recently formed roughly $2\times10^{11} M_{\odot}$ of stars over a comparatively short interval of time, and they may, after $10^9$ years or so, become a galaxy looking very similar to their current companions. And, by the present epoch, this group will almost certainly merge to become a single massive galaxy with G3 dominating its core.

\acknowledgments
We thank the anonymous referee for a thorough review that was helpful in improving our presentation of these results.
The NVAS X-band image shown in Fig.~\ref{vla} was produced as part of the NRAO VLA Archive Survey, \copyright\ AUI/NRAO. The L-band image was kindly produced for us from VLA archival data by Dr Lorant Sjouwerman. This work is based in part on observations made with the Spitzer Space Telescope, which is operated by the Jet Propulsion Laboratory, California Institute of Technology under a contract with NASA. Support for this work was provided by NASA through an award issued by JPL/Caltech. The Gemini GMOS observations were obtained under program GN-2014B-Q80 and the GNIRS under program GN-2015A-Q97.

\bibliographystyle{apj_hack}
\bibliography{ero_ref}

\end{document}